\documentclass[aps,prl,superscriptaddress,twocolumn,balancelastpage,nofootinbib]{revtex4-2}

\usepackage[colorlinks,bookmarks=false,citecolor=blue,linkcolor=blue,urlcolor=blue]{hyperref}
\usepackage[all]{hypcap}   % let hyperlinks correctly point to figures rather than their captions;

\usepackage{amsmath,amssymb}
\usepackage{graphicx}

\usepackage{verbatim}
\usepackage{color}

\usepackage{placeins}    % for FloatBarrier
\usepackage{flafter}     % Bilder immer nach \figure Befehl
\usepackage{color}

\usepackage[normalem]{ulem}

\newcommand{\HIDDEN}[1]{}

\newcommand{\ue}{\text{e}}
\newcommand{\ui}{\text{i}}
\newcommand{\ud}{\text{d}}

\newcommand{\coord}{x}

\makeatletter
\let\Hy@backout\@gobble
\makeatother

\begin{document}

\title{Classical Drift in the Arnold Web Induces Quantum Delocalization Transition}

\author{Jan Robert Schmidt}
\affiliation{TU Dresden,
 Institute of Theoretical Physics and Center for Dynamics,
 01062 Dresden, Germany}

\author{Arnd B\"acker}
\affiliation{TU Dresden,
 Institute of Theoretical Physics and Center for Dynamics,
 01062 Dresden, Germany}

\author{Roland Ketzmerick}
\affiliation{TU Dresden,
 Institute of Theoretical Physics and Center for Dynamics,
 01062 Dresden, Germany}

\date{\today}

\begin{abstract}
We demonstrate that quantum dynamical localization in the Arnold web of
higher-dimensional Hamiltonian systems is destroyed by an intrinsic
classical drift.
Thus quantum wave packets and eigenstates may explore more of the intricate Arnold web than
previously expected.
Such a drift typically occurs, as resonance channels widen toward a large
chaotic region or toward a junction with other resonance channels.
If this drift is strong enough, we find that dynamical localization is destroyed.
We establish that this drift-induced delocalization transition is universal
and is described by a single transition parameter.
Numerical verification is given using a time-periodically kicked Hamiltonian
with a four-dimensional phase space.
\end{abstract}

\maketitle

%%%%%%%%%%%%%%%%%%%%%%%%%%%%%%%%%%%%%%%%%%%%%%%%%%%%%%%%%%%%%%%%%%%%%%%%%%%%%
% Introduction
%%%%%%%%%%%%%%%%%%%%%%%%%%%%%%%%%%%%%%%%%%%%%%%%%%%%%%%%%%%%%%%%%%%%%%%%%%%%%

%%%%%%%%%%%%%%%%%%%%%%%%%%%%%%%%%%%%%%%%%%%%%%%%%%%%%%%%%%%%%%%%%%%%%%%%%%%%%
\vspace*{0.1cm}
\emph{Introduction.}---%
%%%%%%%%%%%%%%%%%%%%%%%%%%%%%%%%%%%%%%%%%%%%%%%%%%%%%%%%%%%%%%%%%%%%%%%%%%%%%
Chaotic dynamics in higher-dimensional Hamiltonian systems and its quantum
mechanical consequences play an important role in many fields of physics.
This is well established for describing atoms, molecules, and chemical
reactions \cite{TanRicRos2000, ManKes2014, WaaSchWig2008, TodKomKonBerRic2005}.
More recently the dynamics in phase space has also been exploited
in the context of quantum many-body systems
\cite{AleKafPolRig2016,BorIzrSanZel2016}.
For example for Bose-Hubbard systems semiclassical methods allow one to explain
many phenomena \cite{RicUrbTom2022,VanBaeKetSch2022,HumRicSch2023},
including spectral statistics, entanglement, and many-body quantum scars.
For current implementations of quantum computers the destructive role of
classical chaos has been demonstrated
\cite{BerVarTreAltDiV2022,BoeBerDiVTreAlt2023:p}.

The classical phase space of such systems is typically governed by the presence
of regions with regular dynamics and regions with chaotic dynamics.
In higher-dimensional systems
all chaotic regions are connected and form a single network, the so-called Arnold
web~\cite{LicLie1992}.
The structure of this network is governed by resonance channels,
characterized by regular dynamics fulfilling
resonant frequency conditions.
The chaotic layers of the resonance channels allow for transport
in the Arnold web by so-called
Arnold diffusion~\cite{Arn1964,Chi1979,LicLie1992,Loc1999,Cin2002,GelTur2017}.
Thus, even in predominantly regular phase-space regions,
the Arnold web makes it
possible to get arbitrarily close to any point in phase space.

In the context of quantum mechanics
an essential question is, how eigenstates and wave
packets explore the intricate Arnold web.
Obviously, the chaotic layer of a resonance channel must be wide enough to
accommodate a wave packet with minimal uncertainty.
But there is another restriction caused by the important phenomenon of dynamical
localization~\cite{CasChiIzrFor1979, CasChi1995:Collection}.
It leads to quantum localization in the presence of classical chaotic diffusion.
For a resonance channel this was demonstrated in a 4D map~\cite{DemIzrMal2002a, DemIzrMal2002b}.
So one expects that a quantum wave packet cannot explore a
resonance channel of the Arnold web.

\begin{figure}
	\includegraphics{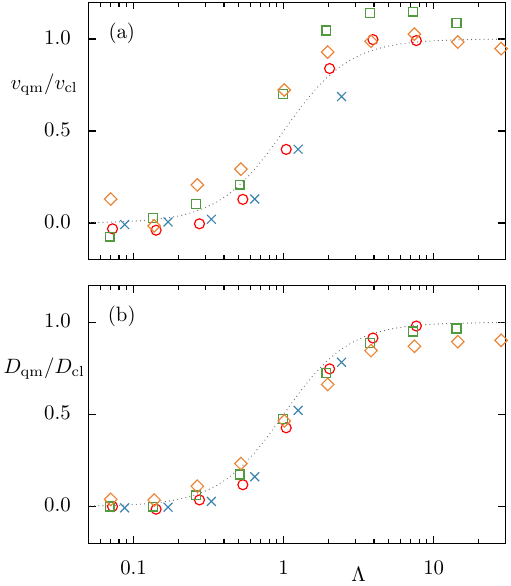}
	\caption{Universal transition from dynamical localization to
        (a) quantum drift with velocity $v_\text{qm}$ and
        (b) quantum diffusion with diffusion coefficient $D_\text{qm}$
        scaled by classical values $v_\text{cl}$ and $D_\text{cl}$, respectively.
        The transition is characterized by the universal transition
        parameter $\Lambda$, Eqs.~\eqref{eq:def_Lambda} and \eqref{eq:Lambda_coord}.
        The function $\Lambda^2/(1+\Lambda^2)$ serves
        as a guide to the eye (dotted line).
        The wave packet dynamics is studied for the kicked Hamiltonian,
        Eq.~\eqref{eq:hamiltonian}, with parameters and symbols described in
        the text.
	}%
	\label{fig:transitions}
\end{figure}

In this paper, however, we demonstrate that dynamical localization is destroyed by a drift in
the Arnold web, if the drift is strong enough.
Such a drift generically occurs when resonance channels widen toward the
chaotic region or toward junctions of resonance channels.
We give a universal description of this drift-induced delocalization transition,
see Fig.~\ref{fig:transitions}.
To this end we introduce a transition parameter,
which depends on the classical drift velocity,
the available chaotic phase-space volume,
and the size of a Planck cell.
We demonstrate the drift-induced delocalization transition using a 4D symplectic map.
Thus quantum wave packets may explore more of the Arnold web
than previously expected.

%%%%%%%%%%%%%%%%%%%%%%%%%%%%%%%%%%%%%%%%%%%%%%%%%%%%%%%%%%%%%%%%%%%%%%%%%%%%%
\vspace*{0.1cm}
\emph{Delocalization transition.}---%
%%%%%%%%%%%%%%%%%%%%%%%%%%%%%%%%%%%%%%%%%%%%%%%%%%%%%%%%%%%%%%%%%%%%%%%%%%%%%
Dynamical localization is a fundamental
phenomenon in quantum chaos if the corresponding classical systems shows
diffusion along a one-dimensional coordinate~\cite{CasChiIzrFor1979, CasChi1995:Collection}.
One finds that the localization length is proportional to the classical diffusion coefficient.
The time evolution of a quantum mechanical wave packet mimics classical
diffusion up to the break time $t^\ast$ at which it localizes with localization
length $\lambda$.

Dynamical localization can be destroyed by various mechanisms, e.g.\
(i) noise induced by coupling to an
environment~\cite{OttAntHan1984,LeiWol1997b,MilSteOskRai2000,GraKol1996,RinSzrGarDel2000,GadReeKriSch2013},
(ii) diffusion in higher dimensions~\cite{AdaTodIke1988, PauBae2020},
or (iii) many-body
interactions~\cite{GliBodFla2011,LelRanDeDelGar2020}, which
recently has been studied experimentally in ultracold
gases~\cite{CaoSajMasSimTanNolShiKonGalWel2022,SeeMcCTanSuLuoZhaGup2022}.

Here we address the general question of dynamical localization in the presence of
an intrinsic classical drift,
which occurs in addition to classical chaotic diffusion.
One intuitively expects that
quantum localization might be destroyed and quantum transport made possible again.
But how strong has the drift to be?

We propose that there is a universal transition from localization to delocalization in
the presence of a classical drift with velocity $v_\mathrm{cl}$.
Without drift, a wave packet localizes at the break time $t^\ast$ with
localization length $\lambda$.
With drift, a wave packet has at this time $t^\ast$ shifted by a distance $|v_\text{cl}| \,
t^\ast$.
If this distance is larger than $\lambda$, no localization with $\lambda$ is
possible.
This motivates the definition of a transition parameter $\Lambda$, which is
given by the ratio of drift distance to localization length in the case of no
drift,
\begin{equation}
    \Lambda = \frac{|v_\text{cl}| \, t^\ast}{\lambda}.
    \label{eq:def_Lambda}
\end{equation}
We expect that localization is destroyed if $\Lambda \gg 1$,
while it should still occur if $\Lambda \ll 1$.

\begin{figure}
	\includegraphics[trim=2.0cm 0.5cm 1.9cm 1.72cm, clip]{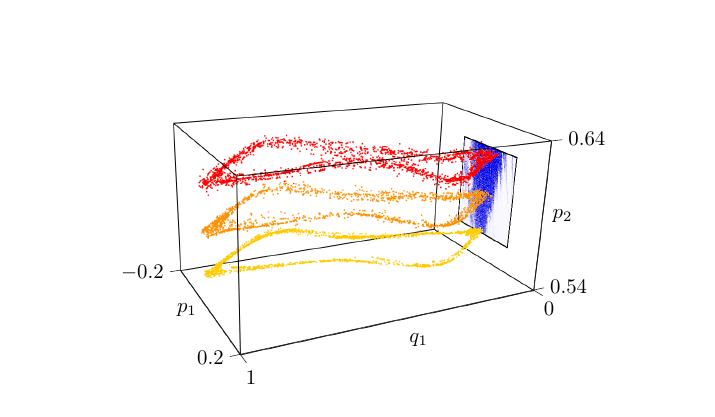}
	% trim=left bottom right top
	\caption{
		3D phase-space slice of the 4D map, Eq.~\eqref{eq:fourD_map},  at $q_2 = 0.5$.
		Three ensembles with initial conditions started at different
		positions $p_2$ near $(q_1, p_1) = (0,0)$ in the resonance channel
		are shown for 200 iterations.
		They spread out uniformly in the chaotic region around the resonance
		and slightly diffuse along $p_2$.
		For larger $p_2$ the chaotic region is wider, as can be also seen
		on the $q_1=0$ plane in which
		the fast Lyapunov indicator from Fig.~\ref{fig:classical_properties}(a)
        is shown.
		}%
	\label{fig:model_visualization}
\end{figure}

\begin{figure}
	\includegraphics{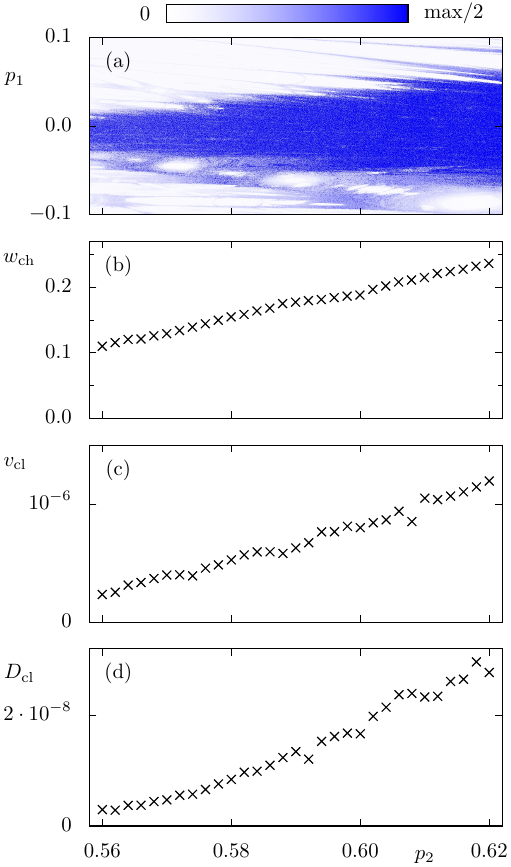}
	\caption{Classical properties of the 4D map, Eq.~\eqref{eq:fourD_map},
		depending on coordinate $x=p_2$:
		(a) Fast Lyapunov Indicator on the plane $q_1 = 0$, $q_2 = 0.5$
		visualizing the increasing cross-sectional volume of the resonance channel,
		(b) chaotic 3D phase-space volume $w_\text{ch}$ of the resonance channel,
		(c) drift velocity $v_\text{cl}$, and
		(d) diffusion coefficient $D_\text{cl}$.
		}%
	\label{fig:classical_properties}
\end{figure}

\begin{figure}
    \includegraphics{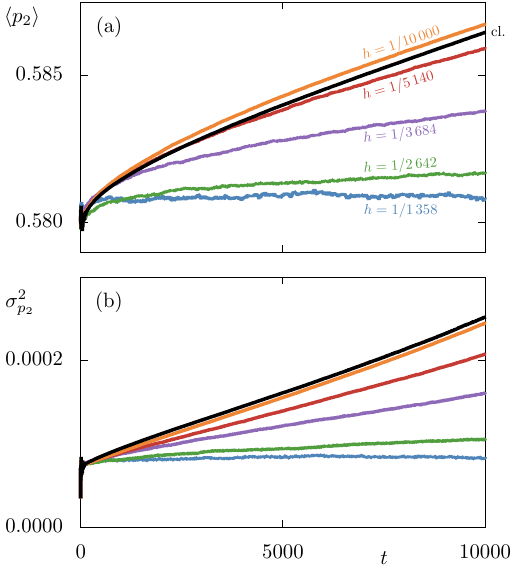}
    \caption{Time dependence of (a) mean value $\langle p_2 \rangle$ and
    	(b) variance $\sigma^2_{p_2}$ along $p_2$ for
        a quantum wave packet for various Planck constants $h$
        (light colors),
        showing a transition from localization to mimicking
        a classical ensemble of initial conditions (black).
        Temporal fluctuations are reduced by convolution with
        a Gaussian of width 1 in $t$.
    }%
    \label{fig:drift_diffusion_measurement}
\end{figure}

Before we provide details of the delocalization transition let us give a
visual overview.
The universal transition at $\Lambda = 1$ is demonstrated in Fig.~\ref{fig:transitions} for
an example system, a kicked Hamiltonian, Eq.~\eqref{eq:hamiltonian}, leading to
a 4D symplectic map.
It has a dominant resonance channel, see Fig.~\ref{fig:model_visualization},
along which classical chaotic diffusion and drift takes place, see
Fig.~\ref{fig:classical_properties}.
The quantum dynamics of a wave packet in this resonance channel is
characterized by a quantum drift velocity and a quantum diffusion coefficient,
see Fig.~\ref{fig:drift_diffusion_measurement}.

%%%%%%%%%%%%%%%%%%%%%%%%%%%%%%%%%%%%%%%%%%%%%%%%%%%%%%%%%%%%%%%%%%%%%%%%%%%%%
\vspace*{0.1cm}
\emph{Transition parameter.}---%
%%%%%%%%%%%%%%%%%%%%%%%%%%%%%%%%%%%%%%%%%%%%%%%%%%%%%%%%%%%%%%%%%%%%%%%%%%%%%
We show for a $2f$-dimensional symplectic map
that the transition parameter $\Lambda$, Eq.~\eqref{eq:def_Lambda},
can be more conveniently expressed in terms of properties of the classical system
and the size $h^f$ of a Planck cell by
\begin{align}
	\Lambda(\coord) = \frac{|v_\text{cl}(\coord)| \; \tau \; w_\text{ch}(\coord)}{h^f}
	\; .
    \label{eq:Lambda_coord}
\end{align}
Here $\coord$ is a local phase-space coordinate along the resonance channel,
$w_\text{ch}(\coord)$ is the cross-sectional volume of the chaotic region
of the resonance channel at $\coord$,
i.e.\ a ($2f - 1$)-dimensional phase-space volume,
and $\tau$ is the time period of the map.
Typically, the drift velocity $v_\text{cl}$
and the volume $w_\text{ch}$
will depend on the coordinate $\coord$ along the resonance channel
and thus so will the transition parameter $\Lambda$.
Note that for $f \ge 2$ there are resonance channels with
1 up to $f-1$ resonance conditions
and we study the maximal case, where the resonance channel
is extended in one dimension.
We expect that these one-dimensional resonance channels have the largest impact on transport.

Equation~\eqref{eq:Lambda_coord} gives $\Lambda$ as
the chaotic $2f$-dimensional phase-space volume explored due to the drift
within one time period of the map
compared to a Planck cell.
Let us remark, that such a ratio of a phase-space volume
to the size of the Planck cell,
also appears in the transport across partial barriers in
phase space~\cite{KayMeiPer1984a, KayMeiPer1984b, Mei1992, Mei2015,
FirBaeKet2023, MicBaeKetStoTom2012, KoeBaeKet2015, StoBaeKet2023:p}.

We can get from Eq.~\eqref{eq:def_Lambda} to
Eq.~\eqref{eq:Lambda_coord} by using the Siberian
argument~\cite{ChiIzrShe1981} generalized to a
one-dimensional resonance channel of a
$2f$-dimensional symplectic map:
We assume that all eigenfunctions in
the chaotic region of
the resonance channel localize with the same
localization length $\lambda$. Then the number of states excited by a
wave packet in the resonance channel is given by
\begin{align}
    \mathcal{N} = \frac{\lambda \; w_\text{ch}}{h^f}
    \; ,
    \label{eq:excited_states}
\end{align}
i.e.\ the chaotic phase-space volume
of the resonance channel
within a localization length
divided by the size of a Planck cell.
The quasienergies $\varepsilon \in [0, \hbar\omega]$
with $\omega = 2 \pi / \tau$
thus have an effective mean level spacing
$\Delta \varepsilon = \hbar \omega / \mathcal{N}$.
This defines an effective Heisenberg time $h / \Delta \varepsilon$,
which equals the break time,
\begin{equation}
    t^\ast = h / \Delta \varepsilon = \mathcal{N} \; \tau
    \; .
    \label{eq:heisenberg_time}
\end{equation}
Combining Eqs.~\eqref{eq:heisenberg_time} and \eqref{eq:excited_states}
gives the ratio $t^\ast / \lambda$,
which leads from Eq.~\eqref{eq:def_Lambda}
to Eq.~\eqref{eq:Lambda_coord}, where the $\coord$ dependence is added.
Note that the localization length is
$\lambda = 2 D_\text{cl} \, w_\text{ch} / h^f$,
which follows from Eqs.~\eqref{eq:heisenberg_time} and \eqref{eq:excited_states}
and by assuming that the variance of classical diffusion with
diffusion coefficient $D_\text{cl}$ at the break time
equals the squared localization length,
$2 D_\text{cl} t^\ast = \lambda^2$.

%%%%%%%%%%%%%%%%%%%%%%%%%%%%%%%%%%%%%%%%%%%%%%%%%%%%%%%%%%%%%%%%%%%%%%%%%%%%%
\vspace*{0.1cm}
\emph{Hamiltonian with large resonance channel.}---%
%%%%%%%%%%%%%%%%%%%%%%%%%%%%%%%%%%%%%%%%%%%%%%%%%%%%%%%%%%%%%%%%%%%%%%%%%%%%%
The Arnold web occurs in Hamiltonian systems
with at least three degrees of freedom or with at least two degrees
of freedom under time-periodic driving.
We concentrate on the lowest dimensional time-periodic case ($f=2$) and a
single resonance channel of the Arnold web.
In order to study the proposed delocalization transition we
need to vary the effective Planck cell over
a sufficiently large range.
This range is quantum mechanically limited
at its lower end due to increasing numerical effort
and at its upper end
by a Planck cell which is still small enough
to fit into the chaotic layer of the resonance channel.

We are able to fulfill these considerations by engineering a Hamiltonian
with a large resonance channel having a large chaotic layer.
Additionally, the cross-sectional volume of the chaotic layer varies by
construction, which induces classically a drift in addition to chaotic
diffusion, as demonstrated below.
We expect, that in this way the generic features of a resonance channel
widening toward the chaotic region or a resonance junction are considered.
The time-periodically kicked Hamiltonian,
based on the coupled standard map~\cite{Fro1972},
is given (in dimensionless units) by
\begin{align}
	H = \, &\frac{p_1^2}{2} + \frac{p_2^2}{2} +  \frac{k(p_2)}{4 \pi^2}
	\cos{(2 \pi q_1)} \sum_{n\, \in \, \mathbb{Z}} \delta(t - n)
	\nonumber \\
	&+ \frac{\xi}{4 \pi^2} \cos{\left(2 \pi [q_1 + q_2]\right)}
	\sum_{n \, \in \, \mathbb{Z}} \delta[t - (n + \epsilon)].
	\label{eq:hamiltonian}
\end{align}
The kicking strength $k(p_2)$ in the first degree of freedom
depens on $p_2$.
The coupling strength $\xi$ governs the coupling between the two degrees of
freedom.
The corresponding kicking term occurs infinitesimally after the
first kicking term by choosing the limit
$\epsilon \to 0^+$ from above.
A different ordering would give a qualitatively similar map.
Periodic boundary conditions with period $1$ are applied to all four coordinates.

%%%%%%%%%%%%%%%%%%%%%%%%%%%%%%%%%%%%%%%%%%%%%%%%%%%%%%%%%%%%%%%%%%%%%%%%%%%%%
\vspace*{0.1cm}
\emph{Classical phase space, drift, and diffusion.}---%
%%%%%%%%%%%%%%%%%%%%%%%%%%%%%%%%%%%%%%%%%%%%%%%%%%%%%%%%%%%%%%%%%%%%%%%%%%%%%
The kicked Hamiltonian \eqref{eq:hamiltonian} leads classically to a
4D symplectic map,
\begin{subequations}
\label{eq:fourD_map}
\begin{align}
    q_1' &= q_1 + p_1, \\
    q_2' &= q_2 + p_2 + \frac{1}{4 \pi^2} \frac{\ud k(p_2)}{\ud p_2}
            \cos{\left(2 \pi q_1' \right)}, \\
    p_1' &= p_1 +  \frac{k(p_2)}{2 \pi}	\sin{(2 \pi q_1')}
            + \frac{\xi}{2 \pi} \sin{\left(2 \pi [q_1' + q_2']\right)}, \\
    p_2' &= p_2
            + \frac{\xi}{2 \pi} \sin{\left(2 \pi [q_1' + q_2']\right)}
    \; .
\end{align}
\end{subequations}
If the coupling $\xi$ is set to zero, $\xi = 0$,
one has a Cartesian product of dynamics in $(q_1, p_1)$
and integrable rotational dynamics in $(q_2, p_2)$.
Such a product structure has been used, e.g., in Refs.~\cite{EasMeiRob2001,GonDroJun2014,
FirBaeKet2023}.
In particular, Eq.~\eqref{eq:fourD_map} corresponds to a stack of 2D standard
maps in $p_2$ direction with kicking strength $k(p_2)$.
We choose $k(p_2)$ such that it gives
rise to an increasing chaotic layer around the main resonance
with a hyperbolic fixed point at $(q_1, p_1) = (0, 0)$
and an elliptic fixed point at $(0.5, 0)$.
For nonzero but small coupling, $\xi=0.1$,
one has a 4D phase space with the topological features of the product structure.
This gives rise to a widening resonance channel along the
$p_2$ coordinate with slow chaotic diffusion due to the weak coupling $\xi$.
For the local phase-space coordinate $x$ along the resonance channel
appearing in Eq.~\eqref{eq:Lambda_coord} we use $x = p_2$.

Specifically, we choose $k(p_2)$ as a periodic triangular function
approximated by a Fourier expansion ranging from $k(0.5) = \bar{k} -
\Delta k$ to $k(0) = k(1) = \bar{k} + \Delta k$ with $\bar{k} = 0.4$ and $\Delta
k = 0.3$ leading to a widening resonance channel,
$k(p_2) = \bar{k} +
\Delta k \; C_n \sum^{n}_{i=0} \cos{[2 \pi (2i+1) p_2]}/(2i+1)^2$ with $n =
2$ and normalization $C_2 = \frac{225}{259}$.
The parameters $\bar{k}$ and $\Delta k$
are chosen such that the influence of other resonance channels
crossing the main one is minimized,
while the increase of the cross-sectional volume $w_\text{ch}(p_2)$
of the resonance channel governed by chaotic dynamics is as
large as possible.

In Fig.~\ref{fig:model_visualization} we present
a 3D phase-space slice~\cite{RicLanBaeKet2014} at $q_2=0.5$ of the 4D map,
Eq.~\eqref{eq:fourD_map},
for three ensembles with initial conditions started at different positions
$p_2$ along the resonance channel
[near $(q_1, p_1) = (0, 0)$ in the first degree of freedom
and for all $q_2$].
Each ensemble spreads in the first degree of freedom along the chaotic layer of
the resonance while staying close to an almost invariant surface $p_2(q_1, p_1)$,
which is flat in the limit $\xi \rightarrow 0$.
Furthermore, it slowly diffuses in $p_2$ direction away from this surface,
which is most prominently seen for the ensemble with largest $p_2$ values.
The dynamics in $q_2$ is purely rotational with momentum $p_2$.

In Fig.~\ref{fig:classical_properties}(a)
the fast Lyapunov indicator~\cite{FroLegGon1997, FroLeg2000, GuzLegFro2002}
in the plane $q_1 = 0$, $q_2 = 0.5$
visualizes the increasing cross-sectional volume of the chaotic region of the
resonance channel, by measuring the chaoticity of trajectories.
In Fig.~\ref{fig:classical_properties}(b) this is
quantified by the 3D chaotic phase-space volume $w_\text{ch}(p_2)$
of the resonance channel at a given coordinate $p_2$.
It is measured by iterating trajectories started in the chaotic sea for
long times and counting the number of visited phase-space boxes.
Over the considered range in $p_2$
an increase by a factor of more than 2 can be observed.
Note that we restrict the analysis to the range $p_2 \in [0.56, 0.62]$, for
which no large resonance channels cross the main one.

Classical transport along $p_2$ is characterized by the drift
velocity $v_\text{cl}(p_2)$ and the diffusion coefficient $D_\text{cl}(p_2)$.
Both strongly depend on $p_2$,
see Figs.~\ref{fig:classical_properties}(c) and
\ref{fig:classical_properties}(d).
Over the considered range in $p_2$
the drift velocity $v_\text{cl}(p_2)$ increases by a factor of 5
and the diffusion coefficient $D_\text{cl}(p_2)$ by a factor of 10.
They are measured by fitting a linear slope to the increasing
mean value and variance in $p_2$ from ensembles of initial conditions
started at each $p_2$,
see as an example the black lines in
Fig.~\ref{fig:drift_diffusion_measurement}.
The time interval used for the linear fit
depends on the initial $p_2$ and is chosen linearly between
$t \in [6\,000, 10\,000]$ for $p_2 = 0.56$ and
$t \in [500, 1\,000]$ for $p_2 = 0.62$.
This ensures that the trajectories (i) are spread out
in the chaotic region of the almost-invariant surfaces
and (ii) are still close to the initial $p_2$.
We attribute the small fluctuations in
Figs.~\ref{fig:classical_properties}(b)-\ref{fig:classical_properties}(d) to the
complex geometry of the Arnold web.
Note that the origin of the drift along the resonance channel is
related to the increasing cross-sectional volume $w_\text{ch}(p_2)$ of the
channel and the increasing diffusion coefficient $D(p_2)$.
It seems possible to quantitatively relate the $p_2$ dependence of the drift
velocity $v_\mathrm{cl}(p_2)$ to $w_\text{ch}(p_2)$ and $D(p_2)$ in the 4D map
using concepts from stochastic processes~\cite{LanBaeKet2016}, but this is
beyond the scope of this paper.
% A quantitative relation to the drift velocity $v_\text{cl}(p_2)$
% in the 4D map
% seems possible using concepts from stochastic processes~\cite{LanBaeKet2016},
% but is beyond the scope of this paper.

%%%%%%%%%%%%%%%%%%%%%%%%%%%%%%%%%%%%%%%%%%%%%%%%%%%%%%%%%%%%%%%%%%%%%%%%%%%%%
\vspace*{0.1cm}
\emph{Quantum delocalization transition.}---%
%%%%%%%%%%%%%%%%%%%%%%%%%%%%%%%%%%%%%%%%%%%%%%%%%%%%%%%%%%%%%%%%%%%%%%%%%%%%%
In order to study the influence of
the classically observed drift on dynamical localization we study the
time evolution of wave packets for the
kicked Hamiltonian~\eqref{eq:hamiltonian}
for varying effective Planck constants $h$, corresponding to
Hilbert space dimensions $1/h^2$ \cite{BerBalTabVor1979,
ChaShi1986, KeaMezRob1999, DegGra2003b, RivSarAlm2000, Lak2001}.
Its time-evolution operator is given by
\begin{align}
    \label{eq:time_evol_op}
    U =
    \ue^{-\frac{\ui}{\hbar} \frac{\xi}{4 \pi^2} \cos{(2\pi[q_1 + q_2])}}
    \ue^{-\frac{\ui}{\hbar} \frac{k(p_2)}{4 \pi^2} \cos{(2 \pi q_1)}}
    \ue^{-\frac{\ui}{\hbar} (\frac{p_1^2}{2} + \frac{p_2^2}{2} )}.
\end{align}
As initial wave packets we choose a product of a coherent state
centered at $(q_1, p_1) = (0, 0)$ in the first degree of freedom
and of a momentum eigenstate at various $p_2$ in the second degree of freedom,
in agreement with the classical initial ensembles.
In Fig.~\ref{fig:drift_diffusion_measurement}
the time dependence of the wave packet's mean value $\langle p_2 \rangle$
and its variance $\sigma^2_{p_2}$ in $p_2$ direction are shown.
A transition from localization for the largest value of $h$
to drift and diffusion for decreasing $h$ can be observed,
mimicking the classical behavior for the smallest value of $h$.

We determine the quantum drift velocity $v_\text{qm}(p_2)$
and diffusion coefficient $D_\text{qm}(p_2)$
from a linear fit using the same time intervals as for
the classical values.
The ratios with the
classical drift velocity $v_\text{cl}(p_2)$
and diffusion coefficient $D_\text{cl}(p_2)$, respectively,
versus the transition parameter $\Lambda$
are shown in Fig.~\ref{fig:transitions}.
We find a smooth transition from localization to
drift-induced delocalization with increasing $\Lambda$
and centered at $\Lambda=1$.
The transition is universal
as it depends on $\Lambda$, Eq.~\eqref{eq:Lambda_coord}, only.
In particular, it is independent of
the specific system parameters $v_\text{cl}(p_2), D_\text{cl}(p_2)$, and
$w_\text{ch}(p_2)$ for various initial
$p_{2} \in \{0.56, 0.59, 0.6, 0.62\}$
and various $h \in [1/500, 1/10\,000]$.
For the smallest Planck constant $h=10^{-4}$ the dimension
of the Hilbert space is $10^8$,
with the time evolution of the kicked Hamiltonian
made possible by using 2D fast Fourier transforms~\cite{AdaTodIke1988}.

%%%%%%%%%%%%%%%%%%%%%%%%%%%%%%%%%%%%%%%%%%%%%%%%%%%%%%%%%%%%%%%%%%%%%%%%%%%%%
\vspace*{0.1cm}
\emph{Discussion and outlook.}---%
%%%%%%%%%%%%%%%%%%%%%%%%%%%%%%%%%%%%%%%%%%%%%%%%%%%%%%%%%%%%%%%%%%%%%%%%%%%%%
We show that the classical drift along resonance channels destroys dynamical
localization if it is strong enough.
The proposed transition parameter $\Lambda$,
Eqs.~\eqref{eq:def_Lambda} and \eqref{eq:Lambda_coord},
leads to a universal description of this drift-induced delocalization
transition.
An important consequence is that a quantum mechanical wave packet may explore
the Arnold web in larger regions than expected.
Namely, the accessible region of a resonance channel, described by a
one-dimensional local phase-space coordinate $x$ along the resonance channel, is
given by those points where $\Lambda(x) \gtrsim 1$.
For strong enough drift the extent of this accessible region is larger than the
localization length obtained from purely diffusive dynamics.
A direct consequence is that also chaotic eigenstates extend further into the
Arnold web.

A future task is to control eigenstates and wave packets in few-body and
many-body systems by adjusting classical drift or relevant phase-space volume.
Interesting examples include many-site Bose-Hubbard systems and quantum computers
with hundreds of qubits.
Extending the analysis to such high-dimensional systems poses a significant
numerical challenge.
Still, we expect the drift-induced delocalization transition to follow the same
universal properties.

%%%%%%%%%%%%%%%%%%%%%%%%%%%%%%%%%%%%%%%%%%%%%%%%%%%%%%%%%%%%%%%%%%%%%%%%%%%%%
\acknowledgments

We are grateful for discussions with Steffen Lange, Martin Richter, and Jonas
Stöber.
This work was funded by the Deutsche Forschungsgemeinschaft (DFG, German
Research Foundation) No.\ 290128388.

%%%%%%%%%%%%%%%%%%%%%%%%%%%%%%%%%%%%%%%%%%%%%%%%%%%%%%%%%%%%%%%%%%%%%%%%%%%%%


\begin{thebibliography}{10}
\newcommand{\enquote}[1]{``#1''}
\providecommand{\url}[1]{\texttt{#1}}
\providecommand{\urlprefix}{URL }
\providecommand{\eprint}[2][]{\url{#2}}

\bibitem{TanRicRos2000}
G.~Tanner, K.~Richter, and J.-M. Rost, \emph{The theory of two-electron atoms:
  between ground state and complete fragmentation}, Rev.~Mod.~Phys.
  \textbf{72}, 497 (2000).

\bibitem{ManKes2014}
P.~Manikandan and S.~Keshavamurthy, \emph{Dynamical traps lead to the slowing
  down of intramolecular vibrational energy flow},
  Proc.~Natl.~Acad.~Sci.~U.S.A. \textbf{111}, 14354 (2014).

\bibitem{WaaSchWig2008}
H.~Waalkens, R.~Schubert, and S.~Wiggins, \emph{Wigner's dynamical transition
  state theory in phase space: classical and quantum}, Nonlinearity
  \textbf{21}, R1 (2008).

\bibitem{TodKomKonBerRic2005}
M.~Toda, T.~Komatsuzaki, T.~Konishi, R.~S. Berry, and S.~A. Rice (editors)
  \emph{Geometric Structures of Phase Space in Multidimensional Chaos:
  Applications to Chemical Reaction Dynamics in Complex Systems}, volume 130 of
  \emph{Advances in Chemical Physics}, John Wiley \& Sons, Inc., Hoboken, New
  Jersey (2005).

\bibitem{AleKafPolRig2016}
L.~D'Alessio, Y.~Kafri, A.~Polkovnikov, and M.~Rigol, \emph{From quantum chaos
  and eigenstate thermalization to statistical mechanics and thermodynamics},
  Adv.~Phys. \textbf{65}, 239 (2016).

\bibitem{BorIzrSanZel2016}
F.~Borgonovi, F.~M. Izrailev, L.~F. Santos, and V.~G. Zelevinsky, \emph{Quantum
  chaos and thermalization in isolated systems of interacting particles},
  Phys.~Rep. \textbf{626}, 1 (2016).

\bibitem{RicUrbTom2022}
K.~Richter, J.~D. Urbina, and S.~Tomsovic, \emph{Semiclassical roots of
  universality in many-body quantum chaos}, J.~Phys.~A \textbf{55}, 453001
  (2022).

\bibitem{VanBaeKetSch2022}
G.~Vanhaele, A.~B{\"a}cker, R.~Ketzmerick, and P.~Schlagheck, \emph{Creating
  triple-{NOON} states with ultracold atoms via chaos-assisted tunneling},
  Phys.~Rev.~A \textbf{106}, L011301 (2022).

\bibitem{HumRicSch2023}
Q.~Hummel, K.~Richter, and P.~Schlagheck, \emph{Genuine many-body quantum scars
  along unstable modes in {Bose}-{Hubbard} systems}, Phys.~Rev.~Lett.
  \textbf{130}, 250402 (2023).

\bibitem{BerVarTreAltDiV2022}
C.~Berke, E.~Varvelis, S.~Trebst, A.~Altland, and D.~P. DiVincenzo,
  \emph{Transmon platform for quantum computing challenged by chaotic
  fluctuations}, Nat.~Commun. \textbf{13}, 2495 (2022).

\bibitem{BoeBerDiVTreAlt2023:p}
S.-D. B{\"o}rner, C.~Berke, D.~P. DiVincenzo, S.~Trebst, and A.~Altland,
  \emph{Classical chaos in quantum computers}, arXiv:2304.14435 [quant-ph]
  (2023).

\bibitem{LicLie1992}
A.~J. Lichtenberg and M.~A. Lieberman, \emph{Regular and {C}haotic {D}ynamics},
  Springer--Verlag, New York, second edition (1992).

\bibitem{Arn1964}
V.~I. Arnol'd, \emph{Instability of dynamical systems with several degrees of
  freedom}, Sov.~Math.~Dokl. \textbf{5}, 581 (1964).

\bibitem{Chi1979}
B.~V. {Chirikov}, \emph{{A universal instability of many-dimensional oscillator
  systems}}, Phys.~Rep. \textbf{52}, 263 (1979).

\bibitem{Loc1999}
P.~Lochak, \emph{Arnold diffusion; {A} compendium of remarks and questions}, in
  C.~Sim{\'o} (editor) \enquote{{H}amiltonian {S}ystems with {T}hree or {M}ore
  {D}egrees of {F}reedom}, volume 533 of \emph{NATO ASI Series: C -
  Mathematical and Physical Sciences}, 168, Kluwer Academic Publishers,
  Dordrecht (1999).

\bibitem{Cin2002}
P.~M. {Cincotta}, \emph{{{Arnold} diffusion: an overview through dynamical
  astronomy}}, New~Astron.~Rev. \textbf{46}, 13 (2002).

\bibitem{GelTur2017}
V.~Gelfreich and D.~Turaev, \emph{{Arnold} diffusion in a priori chaotic
  symplectic maps}, Commun.~Math.~Phys. \textbf{353}, 507 (2017).

\bibitem{CasChiIzrFor1979}
G.~Casati, B.~Chirikov, F.~Izraelev, and J.~Ford, \emph{Stochastic behavior of
  a quantum pendulum under a periodic perturbation}, in G.~Casati and J.~Ford
  (editors) \enquote{Stochastic Behavior in Classical and Quantum Hamiltonian
  Systems}, volume~93 of \emph{Lect.~Notes Phys.}, 334, Springer Berlin /
  Heidelberg, Berlin (1979).

\bibitem{CasChi1995:Collection}
G.~Casati and B.~V. Chirikov (editors), \emph{Quantum chaos: between order and
  disorder}, Cambridge University Press, Cambridge (1995).

\bibitem{DemIzrMal2002a}
V.~Y. Demikhovskii, F.~M. Izrailev, and A.~I. Malyshev, \emph{Manifestation of
  {Arnol'd} diffusion in quantum systems}, Phys.~Rev.~Lett. \textbf{88}, 154101
  (2002).

\bibitem{DemIzrMal2002b}
V.~Y. Demikhovskii, F.~M. Izrailev, and A.~I. Malyshev, \emph{Quantum {Arnol'd}
  diffusion in a simple nonlinear system}, Phys.~Rev.~E \textbf{66}, 036211
  (2002).

\bibitem{OttAntHan1984}
E.~Ott, T.~M. Antonsen, and J.~D. Hanson, \emph{Effect of noise on
  time-dependent quantum chaos}, Phys.~Rev.~Lett. \textbf{53}, 2187 (1984).

\bibitem{LeiWol1997b}
D.~M. Leitner and P.~G. Wolynes, \emph{Intramolecular energy flow in the
  condensed phase: effects of dephasing on localization in the quantum
  stochastic pump model}, Chem.~Phys.~Lett. \textbf{276}, 289  (1997).

\bibitem{MilSteOskRai2000}
V.~Milner, D.~A. Steck, W.~H. Oskay, and M.~G. Raizen, \emph{Recovery of
  classically chaotic behavior in a noise-driven quantum system}, Phys.~Rev.~E
  \textbf{61}, 7223 (2000).

\bibitem{GraKol1996}
R.~Graham and A.~R. Kolovsky, \emph{Dynamical localization for a kicked atom in
  two standing waves}, Phys.~Lett.~A \textbf{222}, 47 (1996).

\bibitem{RinSzrGarDel2000}
J.~Ringot, P.~Szriftgiser, J.~C. Garreau, and D.~Delande, \emph{Experimental
  evidence of dynamical localization and delocalization in a quasiperiodic
  driven system}, Phys.~Rev.~Lett. \textbf{85}, 2741 (2000).

\bibitem{GadReeKriSch2013}
B.~Gadway, J.~Reeves, L.~Krinner, and D.~Schneble, \emph{Evidence for a
  quantum-to-classical transition in a pair of coupled quantum rotors}, Phys.
  Rev. Lett. \textbf{110}, 190401 (2013).

\bibitem{AdaTodIke1988}
S.~Adachi, M.~Toda, and K.~Ikeda, \emph{Quantum-classical correspondence in
  many-dimensional quantum chaos}, Phys.~Rev.~Lett. \textbf{61}, 659 (1988).

\bibitem{PauBae2020}
S.~Paul and A.~B{\"a}cker, \emph{Linear and logarithmic entanglement production
  in an interacting chaotic system}, Phys.~Rev.~E \textbf{102}, 050102(R)
  (2020).

\bibitem{GliBodFla2011}
G.~Gligori{\'c}, J.~D. Bodyfelt, and S.~Flach, \emph{Interactions destroy
  dynamical localization with strong and weak chaos}, EPL \textbf{96}, 30004
  (2011).

\bibitem{LelRanDeDelGar2020}
S.~Lellouch, A.~Ran{\c c}on, S.~De~Bi{\`e}vre, D.~Delande, and J.~C. Garreau,
  \emph{Dynamics of the mean-field-interacting quantum kicked rotor},
  Phys.~Rev.~A \textbf{101}, 043624 (2020).

\bibitem{CaoSajMasSimTanNolShiKonGalWel2022}
A.~Cao, R.~Sajjad, H.~Mas, E.~Q. Simmons, J.~L. Tanlimco,
  E.~{Nolasco-Martinez}, T.~Shimasaki, H.~E. Kondakci, V.~Galitski, and D.~M.
  Weld, \emph{Interaction-driven breakdown of dynamical localization in a
  kicked quantum gas}, Nat.~Phys. \textbf{18}, 1302 (2022).

\bibitem{SeeMcCTanSuLuoZhaGup2022}
J.~H. See~Toh, K.~C. McCormick, X.~Tang, Y.~Su, X.-W. Luo, C.~Zhang, and
  S.~Gupta, \emph{Many-body dynamical delocalization in a kicked
  one-dimensional ultracold gas}, Nat.~Phys. \textbf{18}, 1297 (2022).

\bibitem{KayMeiPer1984a}
R.~S. MacKay, J.~D. Meiss, and I.~C. Percival, \emph{Stochasticity and
  transport in {Hamiltonian} systems}, Phys.~Rev.~Lett. \textbf{52}, 697
  (1984).

\bibitem{KayMeiPer1984b}
R.~S. MacKay, J.~D. Meiss, and I.~C. Percival, \emph{{Transport in
  {H}amiltonian systems}}, Physica~D \textbf{13}, 55 (1984).

\bibitem{Mei1992}
J.~D. Meiss, \emph{Symplectic maps, variational principles, and transport},
  Rev.~Mod.~Phys. \textbf{64}, 795 (1992).

\bibitem{Mei2015}
J.~D. Meiss, \emph{Thirty years of turnstiles and transport}, Chaos
  \textbf{25}, 097602 (2015).

\bibitem{FirBaeKet2023}
M.~Firmbach, A.~B{\"a}cker, and R.~Ketzmerick, \emph{Partial barriers to
  chaotic transport in {4D} symplectic maps}, Chaos \textbf{33}, 013125 (2023).

\bibitem{MicBaeKetStoTom2012}
M.~Michler, A.~B\"acker, R.~Ketzmerick, H.-J. St\"ockmann, and S.~Tomsovic,
  \emph{Universal quantum localizing transition of a partial barrier in a
  chaotic sea}, Phys.~Rev.~Lett. \textbf{109}, 234101 (2012).

\bibitem{KoeBaeKet2015}
M.~J. K\"orber, A.~B\"acker, and R.~Ketzmerick, \emph{Localization of chaotic
  resonance states due to a partial transport barrier}, Phys.~Rev.~Lett.
  \textbf{115}, 254101 (2015).

\bibitem{StoBaeKet2023:p}
J.~St{\"o}ber, A.~B{\"a}cker, and R.~Ketzmerick, \emph{Quantum transport
  through partial barriers in higher-dimensional systems}, arXiv:2308.01162
  [nlin.CD]  (2023).

\bibitem{ChiIzrShe1981}
B.~V. Chirikov, F.~M. Izrailev, and D.~L. Shepelyansky, \emph{Dynamical
  stochasticity in classical and quantum mechanics}, Sov.~Sci.~Rev.~C
  \textbf{2}, 209 (1981).

\bibitem{Fro1972}
C.~{Froeschl\'e}, \emph{Numerical study of a four-dimensional mapping},
  Astron.~Astrophys. \textbf{16}, 172 (1972).

\bibitem{EasMeiRob2001}
R.~W. Easton, J.~D. Meiss, and G.~Roberts, \emph{Drift by coupling to an
  anti-integrable limit}, Physica~D \textbf{156}, 201 (2001).

\bibitem{GonDroJun2014}
F.~Gonzalez, G.~Drotos, and C.~Jung, \emph{The decay of a normally hyperbolic
  invariant manifold to dust in a three degrees of freedom scattering system},
  J.~Phys.~A \textbf{47}, 045101 (2014).

\bibitem{RicLanBaeKet2014}
M.~Richter, S.~Lange, A.~B\"acker, and R.~Ketzmerick, \emph{Visualization and
  comparison of classical structures and quantum states of four-dimensional
  maps}, Phys.~Rev.~E \textbf{89}, 022902 (2014).

\bibitem{FroLegGon1997}
C.~Froeschl\'e, E.~Lega, and R.~Gonczi, \emph{{Fast Lyapunov Indicators}.
  {Application} to asteroidal motion}, Celest.~Mech.~Dyn.~Astron. \textbf{67},
  41 (1997).

\bibitem{FroLeg2000}
C.~Froeschl\'e and E.~Lega, \emph{On the structure of symplectic mappings.
  {The} {Fast Lyapunov Indicator}: a very sensitive tool},
  Celest.~Mech.~Dyn.~Astron. \textbf{78}, 167 (2000).

\bibitem{GuzLegFro2002}
M.~Guzzo, E.~Lega, and C.~{Froeschl{\'e}}, \emph{On the numerical detection of
  the effective stability of chaotic motions in quasi-integrable systems},
  Physica~D \textbf{163}, 1 (2002).

\bibitem{LanBaeKet2016}
S.~Lange, A.~B{\"a}cker, and R.~Ketzmerick, \emph{What is the mechanism of
  power-law distributed {Poincar\'e} recurrences in higher-dimensional
  systems?}, EPL \textbf{116}, 30002 (2016).

\bibitem{BerBalTabVor1979}
M.~V. Berry, N.~L. Balazs, M.~Tabor, and A.~Voros, \emph{Quantum maps},
  Ann.~Phys.~(N.Y.) \textbf{122}, 26 (1979).

\bibitem{ChaShi1986}
S.-J. Chang and K.-J. Shi, \emph{Evolution and exact eigenstates of a resonant
  quantum system}, Phys.~Rev.~A \textbf{34}, 7 (1986).

\bibitem{KeaMezRob1999}
J.~P. Keating, F.~Mezzadri, and J.~M. Robbins, \emph{Quantum boundary
  conditions for torus maps}, Nonlinearity \textbf{12}, 579 (1999).

\bibitem{DegGra2003b}
M.~Degli~Esposti and S.~Graffi, \emph{Mathematical aspects of quantum maps}, in
  M.~Degli~Esposti and S.~Graffi (editors) \enquote{The Mathematical Aspects of
  Quantum Maps}, volume 618 of \emph{Lect.~Notes Phys.}, 49, Springer-Verlag,
  Berlin (2003).

\bibitem{RivSarAlm2000}
A.~Rivas, M.~Saraceno, and A.~{Ozorio de Almeida}, \emph{Quantization of
  multidimensional cat maps}, Nonlinearity \textbf{13}, 341 (2000).

\bibitem{Lak2001}
A.~Lakshminarayan, \emph{Entangling power of quantized chaotic systems},
  Phys.~Rev.~E \textbf{64}, 036207 (2001).

\end{thebibliography}
\end{document}